\documentclass [12pt]{article}
\usepackage {graphicx}
\usepackage[american]{babel}

\begin{document}

\title{A rare and hidden attractor with noise in a biophysical Hodgkin-Huxley-type of model}

\author{N.V. Stankevich}

\maketitle
\begin{center}
\textit{Saint-Petersburg State University, Russia}\\
\textit{University of Jyv\"{a}skyl\"{a}, Finland}\\

\end{center}

\begin{abstract}
{In the present paper the result of numerical simulations of the model based on the Hodgkin-Huxley formalism with bistability between hidden and rare attractors in the presence of noise are studied.\\
\\ }
\end{abstract}


\section{INTRODUCTION}

Multistability and coexistence of dynamical regimes are important properties of cell, neuron, neuron networks and another biological models \cite{a1, a2, a3, a4, a5, a6, a7}. For different applications such property is important for some sense, for instance, for cell multistability can play role of the multypotentiaty, in \cite{a8, a9} authors talk about cancer attractors. If we talk about neuron networks the importance of this property, firstly, connected with interaction in ensembles of such kind systems. It is well-known various phenomena manifested in ensembles of coupled oscillators, as result of their interaction, which can lead to some positive or negative results due their interaction. It has been shown that synchronization plays a key role in the pathogenesis of several neurological diseases, such as Parkinson's disease and essential tremor, i.e. \cite{a10, a11, a12, a13}. For example, it is hypothesized that Parkinsonian symptoms result from asynchronized pacemaker-like activity of a population of many thousands of neurons in the basalganglia \cite{a14, a15}. Another fundamental emergent phenomenon in ensembles of coupled system is oscillation quenching \cite{a16}. The importance of this phenomenon connected with the fact that oscillation suppression or disruption of oscillations is significantly relevant in pathological cases of neuronal disorders such as Alzheimer's and Parkinson's disease. Multistability influents on the dynamics, and in some situation can lead to desynchronization \cite{a17}.

Noise caused by fluctuations at the molecular level is a fundamental part of intracellular and another biological processes. For the systems with multistability the presence of noise can be caused of the switching or another changing of the dynamical behavior as single unit in the network, as whole network or may be formation of some clusters \cite{a18, a19, a20, a21}. In the recent work \cite{a21} the way of controlling of cell state by noise was shown.

In recent paper \cite{a22} was made a big review of so called hidden attractors. In corresponding with \cite{a22} all attractors can be classified into self-excited and hidden. Hidden attractors represent special kind of attractors which are not associated with equilibrium state. It has been shown that multistability is close connected with hidden and rare attractors \cite{a22, a23}. The basins of attraction of the hidden attractors do not contain unstable fixed points and are located far away from such points. If an attractor has very small basin of attraction is called rare attractors. Rare attractors usually is very sensitive to noise and can be destroyed by noise or some perturbations.

In the recent paper \cite{a24} we introduce a modification of well-known Sherman model \cite{a25}, where the coexistence of bursting and silent states (bursting attractor and stable equilibrium point) was observed. In our modification bursting attractor is hidden attractor in respect to stable equilibrium. But the basin of attraction of stable equilibrium is very small. So, we have coexistence of rare and hidden attractors. In the present paper we would like to study the influence of noise on the such kind model. In Sect. 2 we present our modified model and its main parameters and description. In Sect. 3 we consider this model with white noise, and make estimation of switching in the case when one of the coexisting attractors is rare.

\section{Object of studying: modified Sherman}

As basis of the present analysis we will use the following simplified model of the beta-cells suggested by Sherman and Rintzel \cite{a25}:
\begin{equation}
\label{eq:SH_orig}
  \begin{array}{l l l}
    \tau \dot{V} = -I_{Ca} (V) - I_K (V,n) - I_S (V, S),\\
    \tau \dot{n} = \sigma (n_\infty (V) - n),\\
    \tau_S \dot{S} = S_\infty (V) - S.
  \end{array}
\end{equation}
Here, $V$ represents the membrane potential of the cell, $n$ is the fraction of open voltage-gated $K^+$-ion channels. Functions $I_{Ca} (V)$, $I_K (V,n)$ define two intrinsic currents: fast calcium, $I_{Ca}$ and fast potassium, $I_K$:
\begin{equation}
\label{eq:ICa}
I_{Ca} (V) = g_{Ca} m_\infty (V) (V - V_{Ca}),
\end{equation}
\begin{equation}
\label{eq:IK}
I_K (V, n) = g_K n (V - V_K),
\end{equation}
$S$ is the fraction of open voltage-gated $Ca^{2+}$-ion channels, which directly acts on the concentration of $Ca^{2+}$. The third current $I_S (V, S)$ is a $Ca^{2+}$-sensitive slow potassium current, which is directly activated by $Ca^{2+}$.
\begin{equation}
\label{eq:IS}
I_S (V, S) = g_S S (V - V_K),
\end{equation}

The gating variables for $n$ and $S$ are the opening probabilities of the fast and slow potassium currents:
\begin{equation}
\label{eq:OMinf}
\omega_\infty (V) = [1 + exp(\frac{V_\omega - V}{\theta_\omega})]^-1, \omega = m, n, S.
\end{equation}

\begin{table}
\begin{tabular}{|r|l|r|l|r|l|}
  \hline
  $\tau = $ & 0.02 Sec & $\tau_S = $ & 35 Sec & $\sigma = $ & 0.93 \\
  $g_{Ca} = $ & 3.6 & $g_K = $ & 10.0 & $g_S = $ & 4.0 \\
  $V_{Ca} = $ & 25.0 mV & $V_K = $ & -75 mV &  &  \\
  $\theta_m = $ & 12.0 mV & $\theta_n =$  & 5.6 mV & $\theta_S = $ & 10.0 mV \\
  $V_m = $ & -20.0 mV & $V_n = $ & 16.0 mV & $V_S = $ & -35 mV \\
  \hline
\end{tabular}
\caption{Relevant parameters for Sherman model (\ref{eq:SH_orig}).}
\label{tab1}
\end{table}

In the Table~\ref{tab1} there are parameters corresponding to the bursting dynamics of the model (\ref{eq:SH_orig}).

The model (\ref{eq:SH_orig}) has one equilibrium point which undergoes Hobf bifurcation at $V_S = -44.7$ mV. Simultaneously with Hopf bifurcation in model birth bursting attractor. As was shown in \cite{a24} Sherman model (\ref{eq:SH_orig}) can be modified in such way, that unstable equilibrium point will be stabilized through sub-critical Hopf bifurcation and, consequently, there will be co-existence of bursting attractor and stable equilibrium point. To realize stabilization of the equilibrium point  a new voltage-dependent potassium current that varies strongly with the membrane potential right near the equilibrium point was proposed. Introducing of such kind channel can change the sign of the slope of the fast manifold at the equilibrium point (and hence its stability) without affecting the global flow in the model. The form of the new potassium current is the next:
\begin{equation}
\label{eq:IK2}
I_{K2} (V) = g_{K2} p_\infty (V) (V - V_K),
\end{equation}
where $p_\infty (V) = [exp(\frac{V_p - V}{\theta_p})+exp(\frac{V - V_p}{\theta_p})]^{-1}$. All the parameters of the original model (\ref{eq:SH_orig}) can be maintained.
Thus, modified Sherman model has form:
\begin{equation}
\label{eq:SH_modif}
  \begin{array}{l l l}
    \tau \dot{V} = -I_{Ca} (V) - I_K (V,n) - I_{K2} (V) - I_S (V, S),\\
    \tau \dot{n} = \sigma (n_\infty (V) - n),\\
    \tau_S \dot{S} = S_\infty (V) - S.
  \end{array}
\end{equation}
This model has three additional parameters $g_{K2}$, $V_p$, $\theta_p$ which characterize new ion channel. And for this model can coexist bursting attractor and stable equilibrium point for certain values of this parameters, in detail see \cite{a24}. Such kind of modification of the model can be explained, for instance, by the blocking of $K^+$-ion channel, or it inactivation, when voltage on the membrane reach threshold for activation, but the probability of opening channel not equal to 1, and with increasing voltage probability goes to zero \cite{a26, a27}.

\begin{figure}
\centerline{
\includegraphics[scale=0.7]{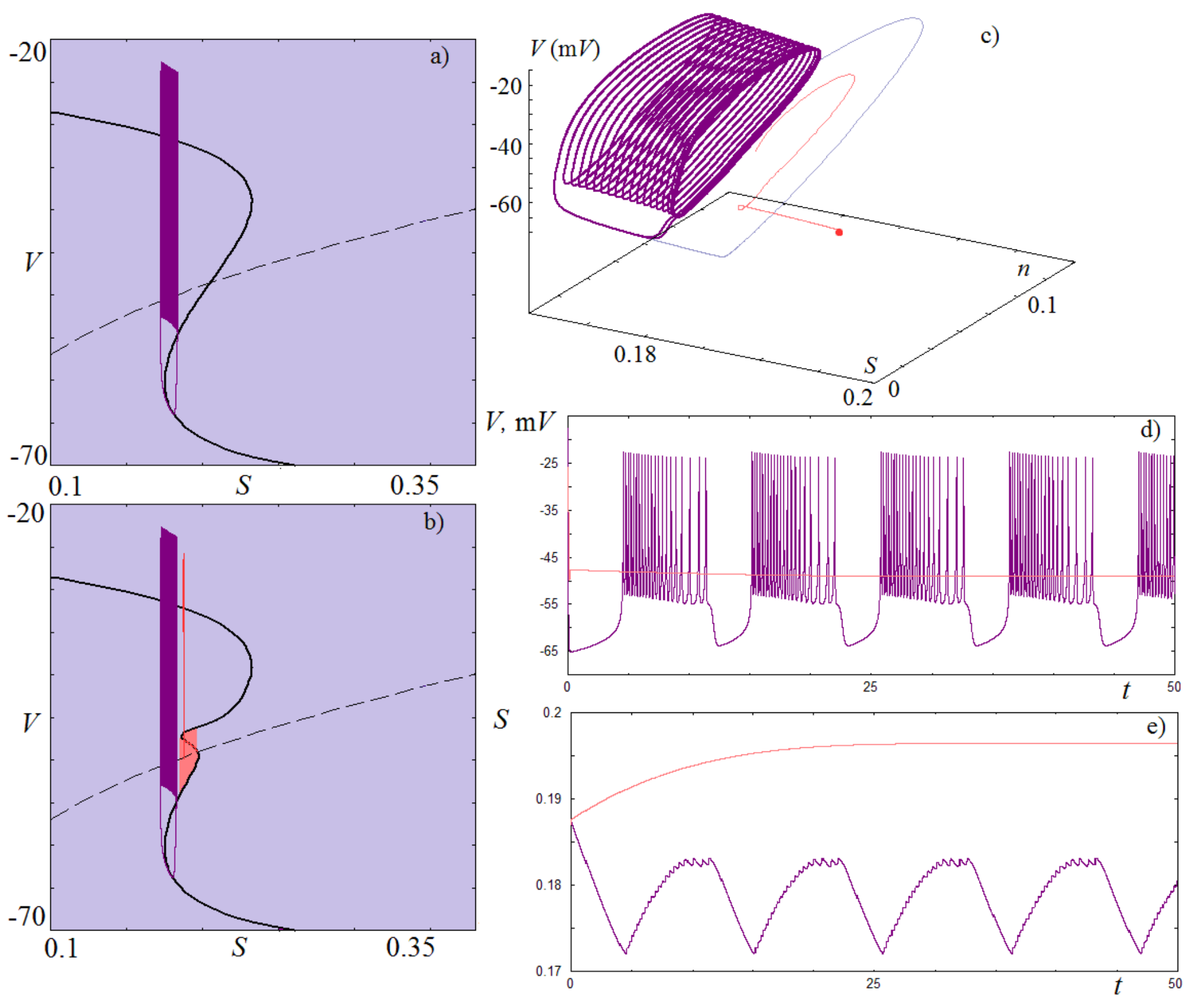}\\}
\caption{a) and b) fast and slow manifolds (black), coexisting dynamical regimes and their basins of attractions for original (purple) and modified (red) models; c) 3D coexisting phase portraits; d) and e) time series of coexisting regimes. $V_p = -47$, $g_{K2} = 0.2$.}
\label{chart1}
\end{figure}

Let us demonstrate coexisting regimes. We fix the new parameters in the following way:
\begin{equation}
\label{eq:SHparam}
g_{K2} = 0.2, V_p = -47, \theta_p = 1.
\end{equation}

For this parameters the model (\ref{eq:SH_modif}) has one stable equilibrium point ($V_0$, $n_0$, $S_0$)=(-49.084, 0.0027105, 0.19648). In Fig.~\ref{chart1} examples of coexisting attractors with (\ref{eq:SHparam}) are presented. In Fig.~\ref{chart1}~a) fast-slow manifolds, projection of bursting attractor and its basin of attraction of the Sherman model (\ref{eq:SH_orig}) are shown, in Fig.~\ref{chart1}~b) fast-slow manifolds, projection of bursting attractor, projection of phase trajectory going to stable equilibrium and its basins of attraction of modified model (\ref{eq:SH_modif}) are shown. Basin of attraction was constructed using Poincar\'{e} section by plane $n = 0.02$, initial condition of the third variable was chosen exact in equilibrium point, $n_0 = 0.00275$. How we can see in Fig.~\ref{chart1}~b) character $S$-form fast manifold of the model has additional bending in which the equilibrium point became stable. This bending is determine the area of attraction of equilibrium stable point. Outside of this area there is area of attraction of bursting attractors. How we can see area of attraction of stable equilibrium point is very small, and we can suggest that this attractor is rare attractor. If we take randomly distributed initial conditions inside cube in phase space, which covers both attractors $V_0 [-65 - -20]$, $n_0 [0 - 0.12]$, $S_0 [0.17-0.2]$, then we get the probability of realization of stable equilibrium point equals 5.5\%, in the case of increasing of this cube this probability will decrease. In Fig.~\ref{chart1}~c) three-dimensional phase portraits of coexisting attractors are shown. The bursting attractor is depicted by purple color, light purple line corresponds to the transition process from initial point located enough far from attractor. The phase trajectory going to the equilibrium point is depicted by red line and dot for the same parameters for what we observe bursting attractor but for another initial conditions. In Fig.~\ref{chart1}~d)-e) time series for coexisting regimes in modified model are shown. So we can see that as result of modification in the system coexistence of stable steady state and bursting attractor is possible.

\section{Bistable model with noise.}

Then we would like to study influence of noise on the dynamics in our modified model (\ref{eq:SH_modif}). Then modified model (\ref{eq:SH_modif}) can be written in the next form:
\begin{equation}
\label{eq:SH_modif_noise}
  \begin{array}{l l l}
    \dot{V} = ( -I_{Ca} (V) - I_K (V,n) - I_{K2} (V) - I_S (V, S)) / \tau + \sqrt{2D} N(t),\\
    \tau \dot{n} = \sigma (n_\infty (V) - n),\\
    \tau_S \dot{S} = S_\infty (V) - S.
  \end{array}
\end{equation}

$N(t)$ is white noise. In our model we have put noise to the first equation because fluctuations is character for membrane voltage, but not to the probability of opening of ion channels.

It is well-known, that bistable systems demonstrate switching at certain level of noise \cite{a19, a20, a21}. And for example, in case when two stable equilibria coexist we can observe alternation of fluctuations near each equilibrium. Let us consider this dynamic in our case.

\begin{figure}
\centerline{
\includegraphics[scale=0.4]{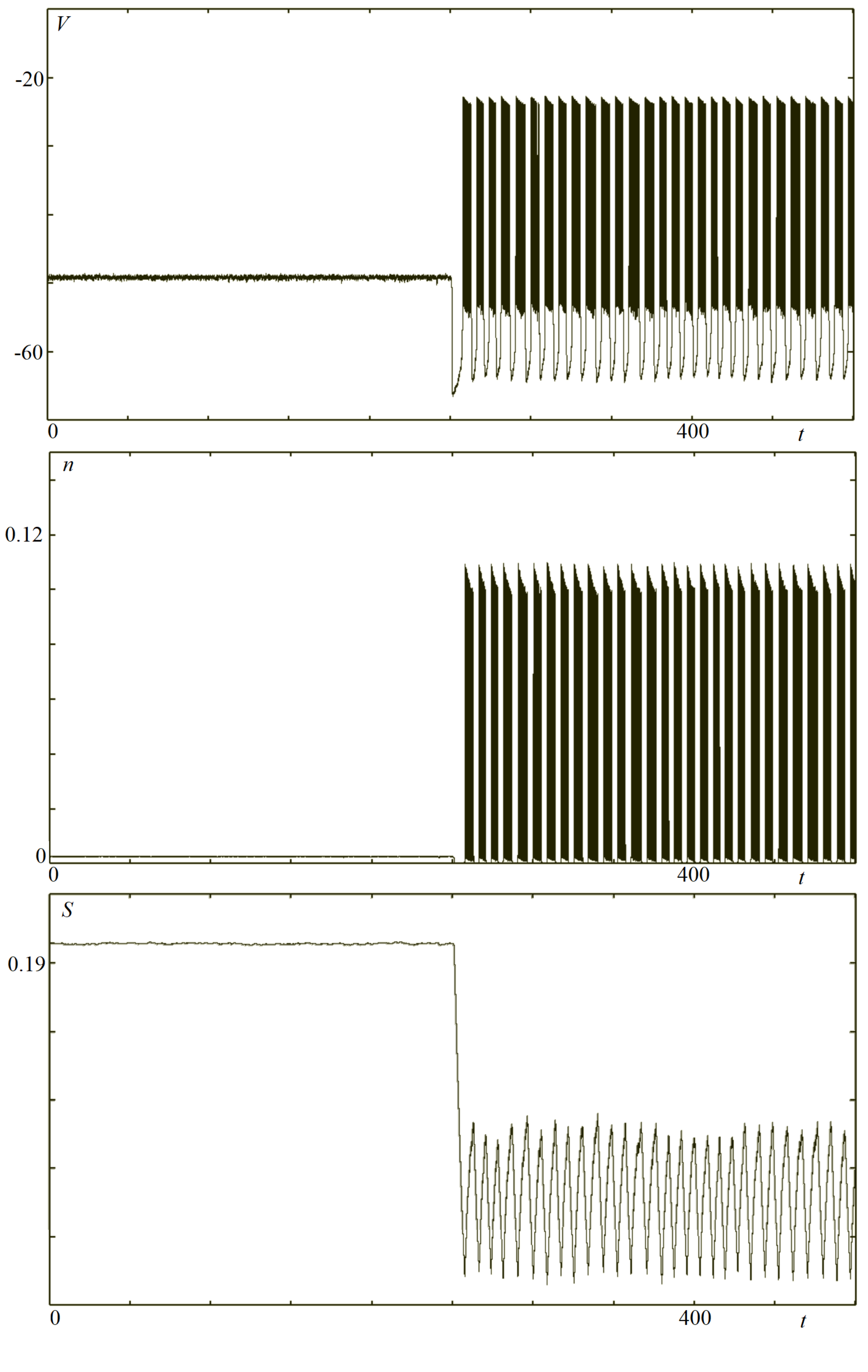}\\}
\caption{Time series for model (\ref{eq:SH_modif_noise}), $D = 0.245$, a) , b), c) correspond to different dynamical variables $V$, $n$, $S$.}
\label{chart2}
\end{figure}

Parameters of the model (\ref{eq:SH_modif}) we fix in corresponding to (\ref{eq:SHparam}). Initial conditions we choose exact in the stable equilibrium point. Then noise with very small intensity induce fluctuations in the vicinity of equilibrium, and it can not reach bursting attractor. If we increase intensity of noise we can realize jump on the bursting attractor ($D\approx0.125$). In Fig.~\ref{chart2} time series for the model (\ref{eq:SH_modif_noise}) at noise $D = 0.245$ are shown.

For time $t = (0-250)$ sec we observe only fluctuations in the vicinity of stable steady state. Then we see jump to bursting attractor. Remark, that at further integration of the model we did not observe jump back to fluctuations in the vicinity of equilibrium. This features connected with:\\
(i) small size of basin of attraction stable equilibrium;\\
(ii) equilibrium point is located enough far from fast-slow nullclines, where is situated bursting attractor;\\
(iii) noise influence on the variable, responding for membrane potential of the cell $V$, but the equilibrium point is distanced in direction of variable $S$, so noise fluctuations can not shift trajectory into basin of attraction equilibrium point.

\begin{figure}
\centerline{
\includegraphics[scale=0.4]{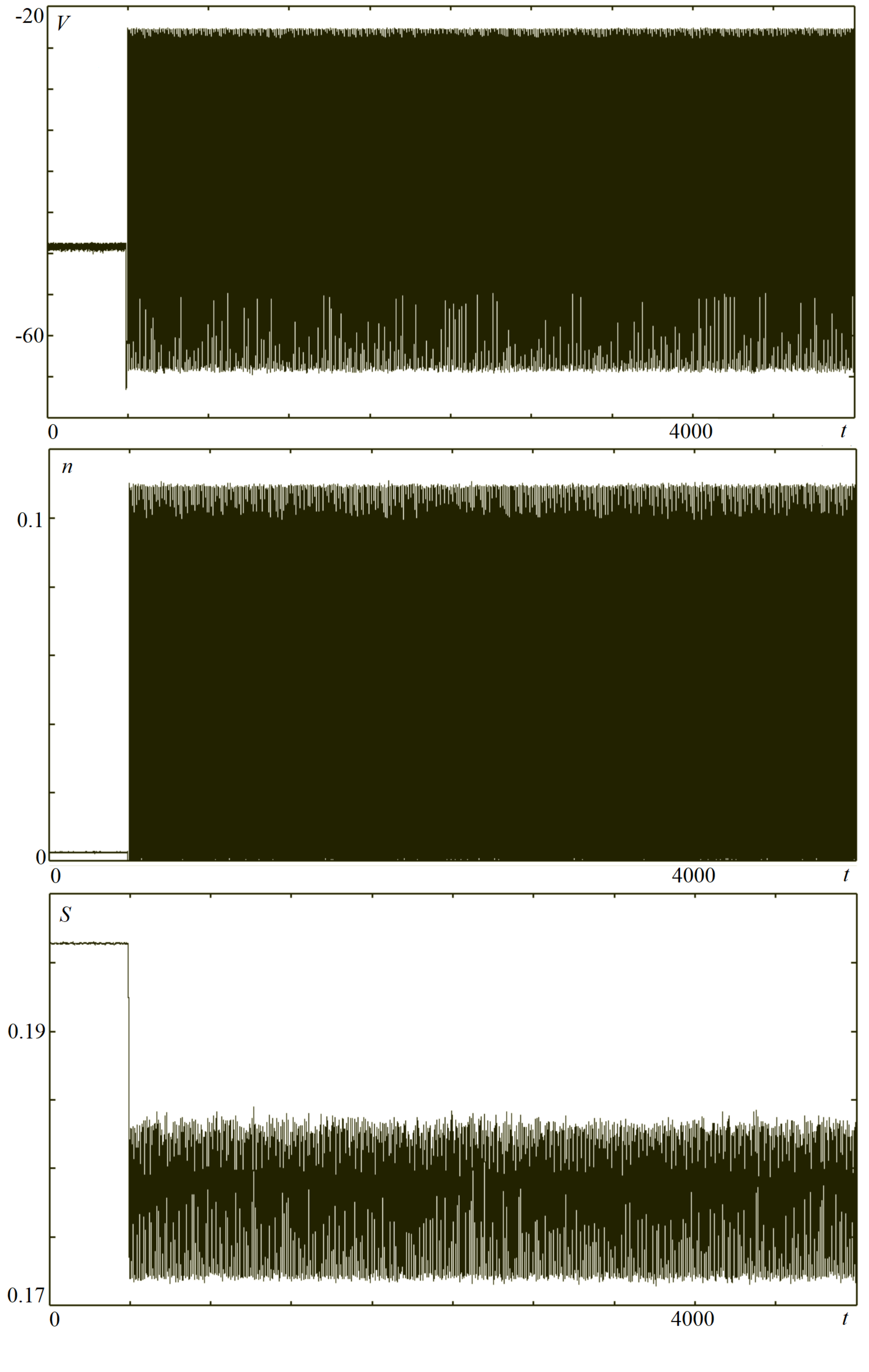}\\}
\caption{Time series for model (\ref{eq:SH_modif_noise}), $D = 0.179$, a) , b), c) correspond to different dynamical variables $V$, $n$, $S$.}
\label{chart3}
\end{figure}

In Fig.~\ref{chart3} another example of time series for less intensity of noise. For less intensity of noise we need more time for jump. But after jump we can not remove our model to the vicinity of equilibrium. So, in this case could not get switching between coexisting regimes.

\section{Conclusion}

Thus in the present paper the result of numerical simulations of the model based on the Hodgkin-Huxley formalism with bistability between hidden and rare attractors are presented. We have shown that in this case we can not observe switching between coexisting dynamical regimes.

\section{Acknowledgments}
This research was leaded in collaboration with A. Koseska and supported by the Russian Science Foundation project 14-21-00041.


\begin{thebibliography}{99}
\bibitem{a1} Pisarchik A.N., Feudel U., Control of multistability, \textit{Physics Reports}, 2014, vol. 540, pp. 167-218.
\bibitem{a2} Heyward, P., Ennis, M., Keller, A., \& Shipley, M. T., Membrane bistability in olfactory bulb mitral cells, \textit{The Journal of Neuroscience}, 2001, vol. 21. pp. 5311-5320.
\bibitem{a3} Loewenstein, Y., Mahon, S., Chadderton, P., Kitamura, K., Sompolinsky, H., Yarom, Y., \& H\"{a}usser, M., Bistability of cerebellar Purkinje cells modulated by sensory stimulation, \textit{Nature neuroscience}, 2005. vol. 8, pp. 202-211.
\bibitem{a4} Izhikevich E.M., Neural excitability, spiking and bursting, \textit{International Journal of Bifurcation and Chaos}, 2000, vol. 10, pp. 1171-1266.
\bibitem{a5} Malashchenko, T., Shilnikov, A., \& Cymbalyuk, G., Six types of multistability in a neuronal model based on slow calcium current, \textit{PloS One}, 2011, vol.6, e21782.
\bibitem{a6} Kelso J.A.S., Multistability and metastability: understanding dynamic coordination in the brain, \textit{Phil. Trans. R. Soc. B}, 2012, vol.367, pp.906-918.
\bibitem{a7} Pomerening, J. R., Sontag, E. D., \& Ferrell, J. E., Building a cell cycle oscillator: hysteresis and bistability in the activation of Cdc2 \textit{Nature cell biology}, 2003, vol.5, pp.346-351.
\bibitem{a8} Huang S., Ernberg I., Kauffman S., Cancer attractors: a system view of tumors from a gene network dynamics and developmental perspective, \textit{Semin Cell Dev Biol}, 2009, vol.20, pp.869-876.
\bibitem{a9} Li Q., Wennborg A., Aurell E., Dekel E., Zou J.-Zh., Xu Y., Huang S., Ernberg I., Dynamics inside the cancer cell attractors reveal cell heterogenety, limits of stability, and escape, \textit{PNAS}, 2016.
\bibitem{a10} P. Tass, M. Rosenblum, J. Weule, J. Kurths, A. Pikovsky, J. Volkmann, A. Schnitzler, H.-J. Freund, Detection of n:m phase locking from noisy data: application to magnetoencephalography, \textit{Phys.Rev.Lett.}, 1998, vol.81, pp.3291-3294.
\bibitem{a11} P.A. Tass, Phase Resetting in Medicine and Biology. Stochastic Modelling and Data Analysis, Springer, Berlin, 1999.
\bibitem{a12} J.Milton, P.Jung, Epilepsyasa Dynamic Disease, Springer, Berlin, 2003.

\bibitem{a13} G.Buzsaki, A.Draguhn, Neuronal oscillations in cortical networks, Science 304 (2004) 1926-1929.
\bibitem{a14} H.Bergman, A.Feingold, A.Nini, A.Raz, H.Slovin, M.Abeles, E.Vaadia, Physiological aspects of information processing in the basalganglia of normal and parkinsonian primates, TrendsNeurosci. 21 (1998) 32-38.
\bibitem{a15} J.Sarnthein, A.Morel, A.vonStein, D.Jeanmonod, Thalamic theta-field potentials and EEG: high thalamocortical coherence in patients with neurogenic pain, epilepsy and movement disorders, Thalamus \& Related Systems 2 (2003) 231-238.
\bibitem{a16} A. Koseska, E. Volkov, J. Kurths. Oscillation quenching mechanisms: Amplitude vs. oscillation death. PhysicsReports, 531 (2013) 173-199.
\bibitem{a17} B. Lysyansky, Y. L. Maistrenko, and P. A. Tass. Coexistence of numerous synchronized and desynchronized states in a model of two phase oscillators coupled with delay. International Journal of Bifurcation and Chaos 18.06 (2008): 1791-1800.
\bibitem{a18} R. Perez-Carrasco, P. Guerrero, J. Briscoe, K. M. Page. Intrinsic Noise Profoundly Alters the Dynamics and Steady State of Morphogen Controlled Bistable Genetic Switches //  PLoS Comput Biol 12(10): e1005154.doi:10.1371/ journal.pcbi.1005154.
\bibitem{a19} A.N. Pisarchik, I.A. Bashkirtseva, and L.B. Ryashko. Controlling bistability in a stochastic perception model. Eur. Phys. J. Special Topics 224, 1477-1484 (2015)
\bibitem{a20} A. Koseska, A. Zaikin, J. Kurths, J. Garc\'{\i}a-Ojalvo. Timing Cellular Decision Making Under Noise via Cell- Cell Communication. PLoS ONE 4(3): e4872.
\bibitem{a21} Wells D.K., Kath W.L., and Motter A.E. Control of Stochastic and Induced Switching in Biophysical Networks, Phys. Rev. X, 5, 031036 (2015).
\bibitem{a22} D. Dudkowski, S. Jafari, T. Kapitaniak, N.V. Kuznetsov, G.A. Leonov, A. Prasad, Hidden attractors in dynamical systems, Physics Reports, Vol. 637, pp. 150 (2016).
\bibitem{a23} S. Brezetskyi, D. Dudkowskia, and T. Kapitaniak. Rare and hidden attractors in Van der Pol-Duffing oscillators. Eur. Phys. J. Special Topics 224, 1459-1467 (2015)
\bibitem{a24} Stankevich N.V., Mosekilde E. Coexistence between bursting and silent states in a biophysical model of Hodgkin-Huxley-type CHAOS (submitted).
\bibitem{a25} Sherman A. Anti-phase, asymmetric and aperiodic oscillations in excitable cells I. Coupled bursters, Bulletin of mathematical biology. 1994. Vol. 56, No 5. P. 811-835.
\bibitem{a26} Litan A., and Langhans S.A. Cancer as a channelopathy: ion channels and pumps in tumor development and progression, Frontiers in cellular neuroscience 9, 86 (2015).
\bibitem{a27} Simms B.A., and Zamponi, G.W. "Neuronal voltage-gated calcium channels: structure, function, and dysfunction", Neuron, 82(1), 24-45 (2014).
\end{thebibliography}
\end{document}